\documentstyle[psfig]{EuroPhys}

\def\etal{{\hbox{{\tenit\ et al.\/}\tenrm :\ }}}

\newif\ifboo \boofalse

\begin{document}
%
%
\euro{42}{6}{655}{1998}
\Date{15 June 1998}
\shorttitle{M. ULMKE \etal ENHANCEMENT OF LONG-RANGE AF ORDER}

\title{Enhancement of Long Range Antiferromagnetic Order by 
Nonmagnetic Impurities in the Hubbard Model}

\author{M.~Ulmke\inst{1}, P.~J.~H.~Denteneer\inst{2}, R.~T.~Scalettar\inst{3},
G.~T.~Zimanyi\inst{3}}

\institute{
\inst{1} Theoretische Physik III, Elektronische Korrelationen und Magnetismus,
Institut f\"ur Physik, Universit\"at Augsburg, D 86135 Augsburg, 
Germany \\
\inst{2} Lorentz Institute for Theoretical Physics, 
University of Leiden, P.O.~Box 9506, 2300 RA Leiden, The Netherlands \\ 
\inst{3} Department of Physics, University of California, Davis, CA 95616, USA}

\rec{31 October 1997}{in final form 4 May 1998}

\pacs{
\Pacs{71}{27$+$a} {Strongly correlated electron systems; heavy fermions}
\Pacs{71}{30$+$h}{Metal-insulator transitions and other electronic transitions}
}

\maketitle

\begin{abstract} 

The two--dimensional Hubbard model with a bimodal distribution of 
on--site interactions, $P(U_{\bf i})=(1-f)\delta(U_{\bf i}-U)
+f\delta(U_{{\bf i}})$,
is studied using finite temperature Quantum Monte Carlo 
and dynamical mean--field theory. 
Long range antiferromagnetic order off
half--filling is {\it stabilized} by the disorder,
due to localization of
the dopants on the $U=0$ sites.
Whereas in the clean model there is a single gap at $n=1$,
for nonzero $f$ we find 
the compressibility and density of states 
exhibit gaps at two separate fillings.

\end{abstract}

\vspace{-10mm}

The Hubbard model exhibits both Mott metal--insu\-lator and magnetic 
phase transitions.  
On the one hand, the interaction $U$
induces a gap at half--filling 
by separating many--body states with doubly
occupied orbitals (upper Hubbard band) from those with holes 
(lower Hubbard band) when $U$ becomes larger
than the non--interacting bandwidth $W$. 
On the other hand, also near half--filling,
there is a tendency to antiferromagnetic (AF) ordering 
of the electron spins.  For interacting fermions,
the issue has recently been raised whether
the Mott transition ever occurs in the absence of
associated symmetry breaking such as magnetic order \cite{GEBHARD}.

The introduction of ``impurity'' sites where $U=0$ in principle
could separate AF order and the Mott
transition.  The Mott gap will be shifted to densities greater than $n=1$
since some sites can be doubly occupied without any on--site
repulsion energy cost.  Meanwhile, it is likely that 
the Fermi--surface instability
responsible for opening the AF (spin density wave) gap
remains at half--filling.
In this paper we will study a model Hamiltonian incorporating this
effect.  Our main conclusions are:  (1)  $U=0$ sites can induce
long range AF order at densities which are spin disordered 
in the clean model, through localization of the doped particles.
(2)  The dependence of the occupation on the
chemical potential and the behavior of the density of states
exhibit a Mott gap, shifted from half--filling, and also
a gap at half--filling resulting from induced AF
order on the $U=0$ sites.

There are a number of experimental systems where the effect
of the introduction of nonmagnetic impurities has been studied.
Examples include doping Zn or Ga for La in La$_{2}$CuO$_{4}$,
where the critical concentration for the destruction of AF order,
$x_c \approx 0.10-0.15$ is considerably larger than for doping
with Sr, which is not isovalent \cite{ZN1},
and also Zn doping in ladder compounds where an AF phase 
is stabilized \cite{ZN2}.
Numerical work on the effect of impurities on  N\'{e}el
order has focused on spin systems \cite{NUMERICS}.
Enhanced local correlations arise from
restrictions on the singlet bond patterns 
due to the defects \cite{MARTINS}.  

We study the electronic lattice Hamiltonian,
\begin{eqnarray}
\hat H &=&-t \sum_{<{\bf ij}>,\sigma} \hat c_{{\bf i}\sigma}^\dagger    
        \hat c^{\phantom \dagger}_{{\bf j}\sigma}
        -\mu \sum_{{\bf i}\sigma} \hat n_{{\bf i}\sigma}
        +\sum_{\bf i} U_{\bf i} (\hat n_{{\bf i}\uparrow}-\frac{1}{2})
                           (\hat n_{{\bf i}\downarrow}-\frac{1}{2}) \; .
\label{model}
\label{eq1}
\end{eqnarray}

\noindent
The kinetic energy is described by hopping  between nearest--neighbor 
lattice sites, $\langle{\bf ij}\rangle$, on a square ($N$=$L \times L$) 
lattice.
The interaction values $U_{\bf i}$ are 
chosen from a bimodal distribution, $U_{\bf i}\in \{0,U\}$,
with probabilities $f$ and $1-f$, respectively.
Upon electron doping the $U=0$ sites will be filled thereby
quenching their local moment and eventually acting 
as magnetically inert impurities.

We use two complementary numerical approaches,
finite dimensional ``determinant'' quantum Monte Carlo (QMC) algorithm 
\cite{Blankenbecler81,Sugiyama86etc}
and the closely related ``dynamical mean field theory'' (DMFT) 
\cite{Metzner89etc,Jarrell92,Georges96}, also employing QMC \cite{Hirsch86}.
The former technique allows an approximation free solution
of the equilibrium thermodynamic properties, but only on
finite lattices and in a somewhat restricted range of parameters.
The latter technique introduces a local approximation to the self--energy, 
which becomes exact in the limit of infinite dimensions,
and allows a more complete exploration of parameter space
in the thermodynamic limit.
DMFT has proven to give reliable results for thermodynamic and dynamic
properties in $d=2$ and $d=3$ \cite{Jarrell92,Georges96},
and has also been applied to study disorder effects on interacting
systems \cite{Dobrosalievic93etc,Ulmke95}.
While for $d=2$ expectation values are averaged over typically tens of 
(static) disorder configurations, in DMFT this average reduces to a 
weighted sum over the two possible values of $U_i$,
equivalent to the coherent--potential approximation \cite{Elliot74}.

Like the original Hubbard model, Hamiltonian (\ref{model}) 
is particle--hole symmetric at half--filling ($n=1$ and $\mu=0$), 
i.e.~it is invariant under the ``staggered'' particle--hole transformation 
$\hat c^\dagger_{{\bf i}\sigma}\rightarrow 
(-1)^{\bf i} \hat c^{\phantom \dagger}_{{\bf i}\sigma}$,
so there is no minus--sign problem at $n=1$
which would preclude simulations for large lattices at low temperatures $T$.
Physically, the particle--hole symmetry corresponds to different chemical 
potentials on the two constituents such that at $\mu=0$ the 
local density expectation values $n_{\bf i}$ are homogeneous, 
i.e.~independent of $U_{\bf i}$ \cite{phsymmetry}.
The choice here of the symmetric limit of the model is identical in
spirit to the focus in the heavy fermion literature on the
`symmetric' Anderson model, in which a similar specific relation
between f-site energy and f interaction strength is assumed to bring
out most clearly the physics of that Hamiltonian.
In the following we will discuss the effects of disorder on (i) AF
order and (ii) the charge gap, presenting for each case first $d=2$ 
data which are then complemented by DMFT results.

\vspace{1.5em}

\begin{figure}[t]
\vspace{-7mm}
\hspace*{32mm}
{\psfig{file=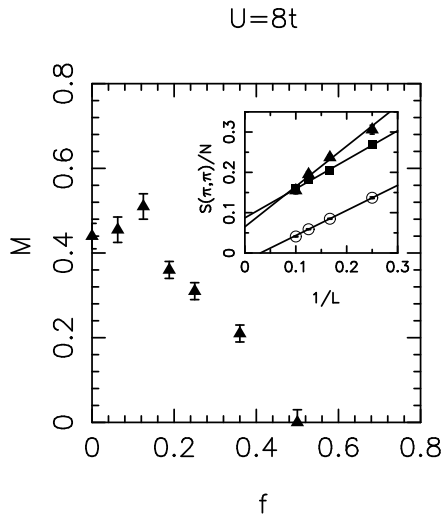,width=4.in,angle=-90}}
\vspace{-5mm}
\caption{Ground state staggered magnetization $M$ as a function of 
fraction of $U=0$~sites, $f$, for $U=8t$, $n=1$. 
Inset: Finite--size scaling of the antiferromagnetic structure factor.
 $f=$0.0, 0.125,
(filled triangles and squares) are ordered.
$f=0.500$ (open circles) is not.
For values of $f$ that do not correspond to an integer number of defects, 
we have interpolated the results for the two bracketing concentrations.
\label{mstag}}
\vspace{-5mm}
\end{figure}

\em AF order. \em  ---
First we consider the case $n=1$ and 
study the effect of an increasing concentration $f$ of $U=0$ sites
on the stability of AF long range order. 
The static AF structure factor $S(\pi,\pi)$ is calculated for 
different lattice sizes $(L\leq 10)$ and temperatures, and averaged
over up to 40 disorder configurations.
For $L\leq 10$, $S(\pi,\pi)$ 
is found to saturate at $T\approx t/8$. From the saturated values
the ground state sublattice magnetization $M$ can be extrapolated using a 
finite--size scaling according to spin wave theory \cite{Huse88}, 
$S(\pi,\pi)/L^2 =M^2/3 + O(1/L)$. 
Scaling plots for different $f$ at $U=8t$ are shown in 
the inset of Fig.~\ref{mstag}. 
For $f\leq 0.36$, $S(\pi,\pi)/L^2$
extrapolates to a finite value in the thermodynamic limit.
For $f=0.5$ the linear scaling with $1/L$ does not hold anymore,
since it leads to negative values of $M^2$,
indicating the absence of AF order in the thermodynamic limit.
Fig.~\ref{mstag} presents the 
extrapolated values of $M$ as a function of $f$.
A spin wave suppression of the staggered magnetization\cite{CPA}.
might be the origin of the weak enhancement of $M$ near $f=0.1$.

\begin{figure}[t]
\vspace*{-10mm}
\hspace*{-10mm}
{\psfig{file=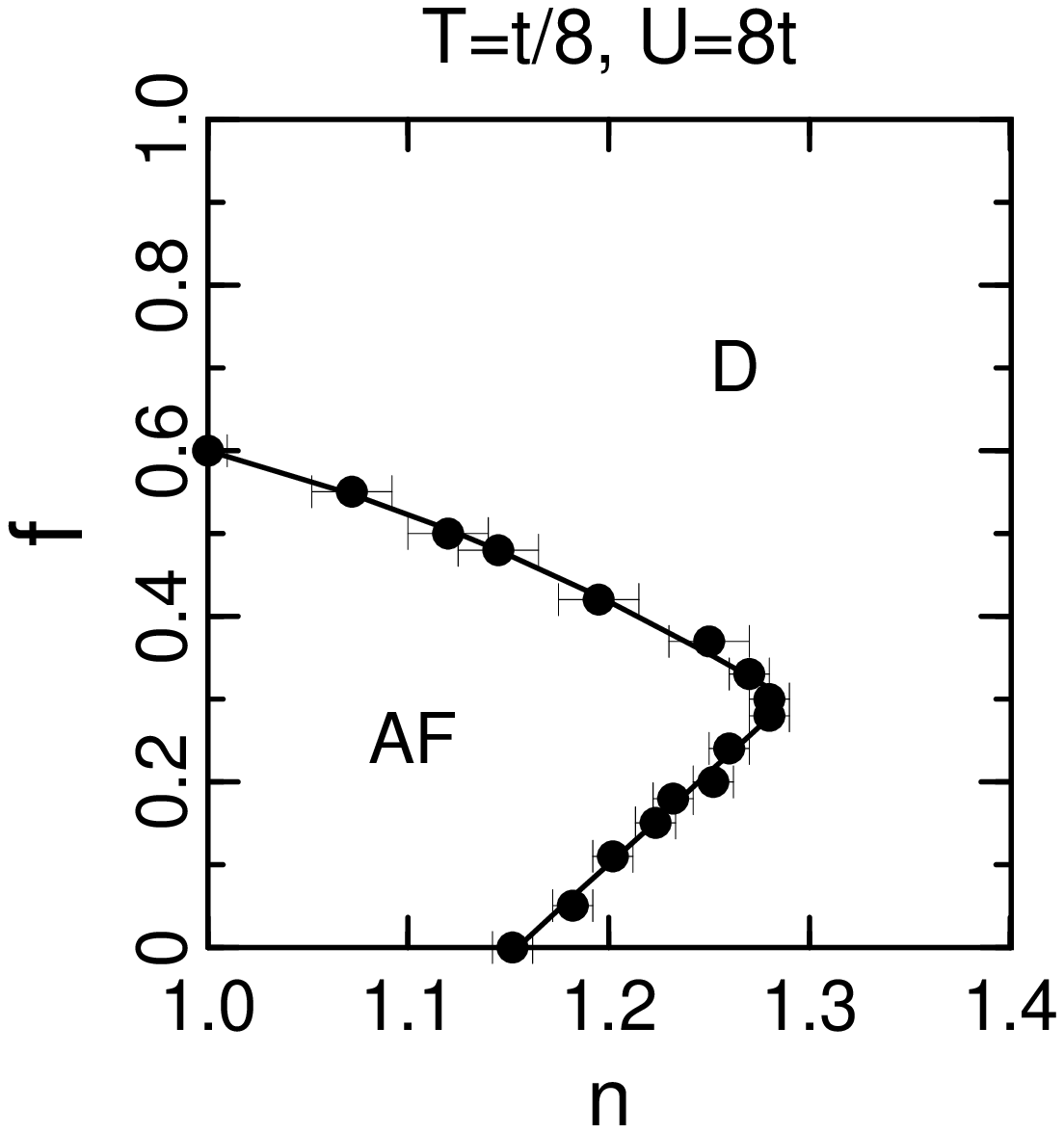,width=4.in,angle=0}}
\vspace{-15mm}
\caption{
Phase diagram in DMFT as a function of filling, $n$, and fraction of
$U=0$ sites, $f$.  For $0<f<0.4$ the antiferromagnetic (AF) region
around half--filling is enhanced relative to $f=0$.  At
fixed density $n > 1$, the introduction of $U=0$ impurities 
localizes the mobile dopants and restores long range order.
\label{phased}}
\vspace{-86.5mm}
\hspace*{65mm}
{\psfig{file=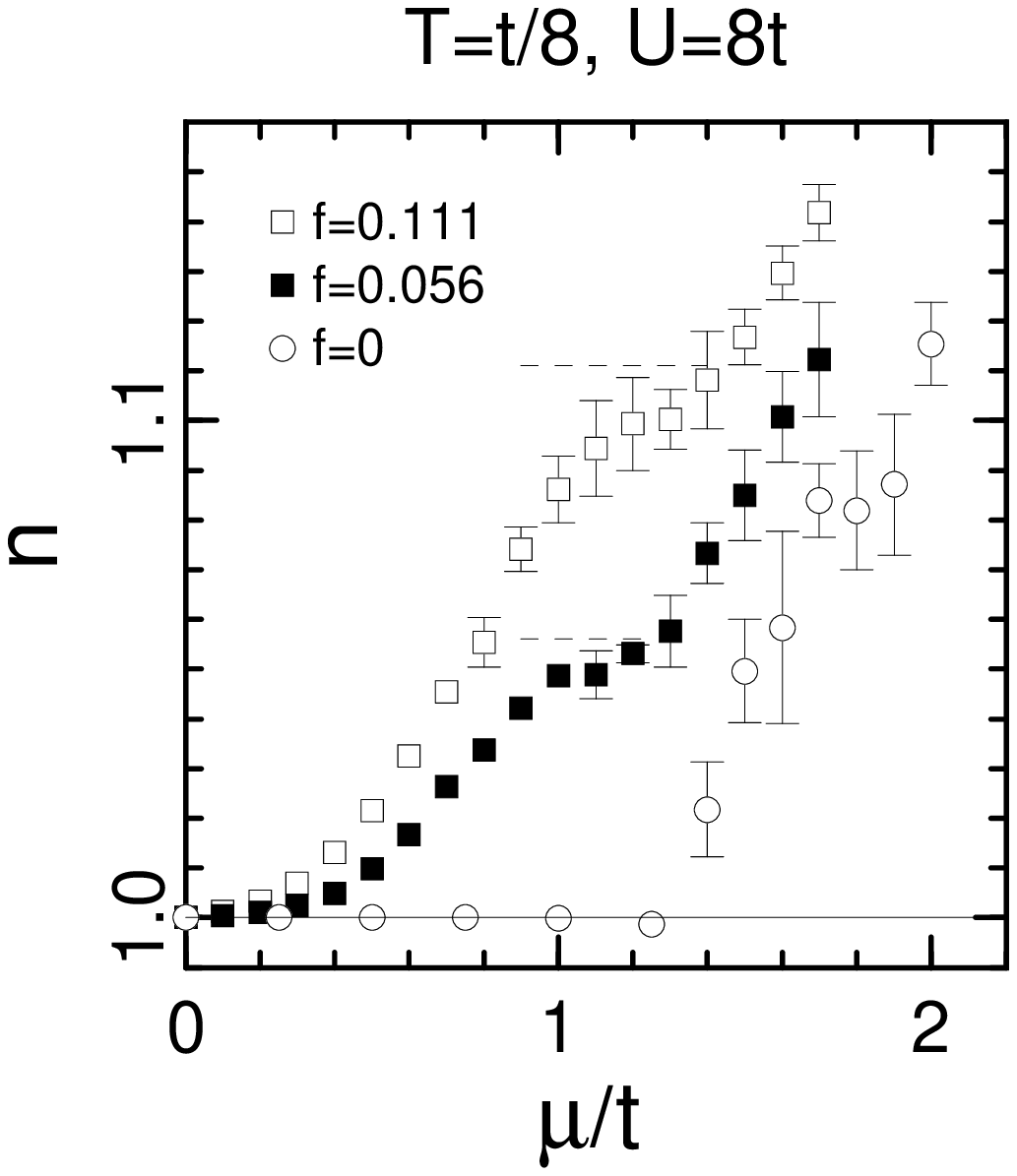,width=4.in}}
\vspace{0mm}
\caption{
Density $n$ vs.~chemical potential $\mu$ for 
different values of $f$ on a $6\times 6$ lattice with $U=8t$.
Error bars are within the symbol size when not shown [18].
The Mott plateau is shifted away from half--filling.
Dashed lines indicate the values $1+f$ (corresponding to
complete occupation of $U=0$ sites).
\label{nvsmud2}}
\vspace{-2mm}
\end{figure}

The $U=0$ impurity sites can induce AF order
at a doping for which the clean model is spin disordered.
At finite doping the minus--sign problem prevents the simulations
necessary for the finite--size analysis. 
We therefore employ DMFT to calculate the phase diagram 
in the density--disorder plane, shown in Fig.~\ref{phased} for $T=t/8$.
The phase boundary is obtained both from the vanishing
of $M$ in the AF phase and from the divergence of the staggered
susceptibility in the spin disordered (D) phase.
The region of AF long range order extends to a larger doping
for finite $f$; the presence of impurity sites
stabilizes order.  In the clean model, doping away from half--filling
introduces extra particles which are mobile and hence
especially effective at disturbing correlations over the entire lattice.
$U=0$ defects provide localizing sites which are energetically favorable
for these extra particles. 
Thus, while doping destroys moments
locally, it has a much smaller effect on long range correlations.
In two dimensions, the critical doping value
for the clean $f=0$ model is smaller than in DMFT and probably zero
because of competing charge or incommensurate magnetic instabilities.
However, we expect that the
enhancement of the AF phase by nonzero $f$ will also be present in $d=2$,
since the underlying mechanism, the defect--induced localization, 
is observed in $d=2$, too.

\vspace{1.em}
\em Charge gap. \em ---
In the extreme strong coupling limit, $t=0$, a plot of the
density $n$ as a function of chemical potential $\mu$
exhibits plateaus at $n=1 \pm f$:
$n$ is zero until 
$\mu=-U/2$, at which point $n$ will jump to $n=1-f$ (singly
occupying sites which have nonzero $U$.)  When $\mu$ goes through
zero, $n$ will jump to $n=1-f + 2f = 1+f$ (singly occupied nonzero
$U$ sites plus doubly occupied $U=0$ sites).  Finally, at $\mu=+U/2$, the
density jumps to $n=2$ (all sites doubly occupied). 
When $t$ is turned on these sharp steps will be rounded by quantum
fluctuations, but we might expect
the compressibility $\kappa = \partial n / \partial \mu$ to remain large
at $n=1$ and remain small at $n=1+f$.
In order to explore precisely the effect of nonzero hopping 
the dependence of $n$ on the chemical potential
is calculated using $d=2$ QMC.
Fig.~\ref{nvsmud2} shows results for a $6\times 6-$lattice at
$T=t/8$ and $U=8t$ \cite{ERRORBARS}.
At $f=0$ there is a wide charge gap, visible in the broad plateau
at $n=1$ starting at $\mu=0$.
For $f>0$ the ``Mott'' gap occurs close to the expected density
of $n=1+f$ (dashed lines).  It terminates at a chemical potential
close to the $\mu_{c}$ for $f=0$.  That is, the chemical
potential to force double occupation of the non--zero $U$ sites
is unaffected by the presence of the $U=0$ sites.
In addition, however, a second plateau remains at half--filling,
an effect not predicted by the $t=0$ analysis.

\begin{figure}
\vspace*{5mm}
\hspace*{-10mm}
{\psfig{file=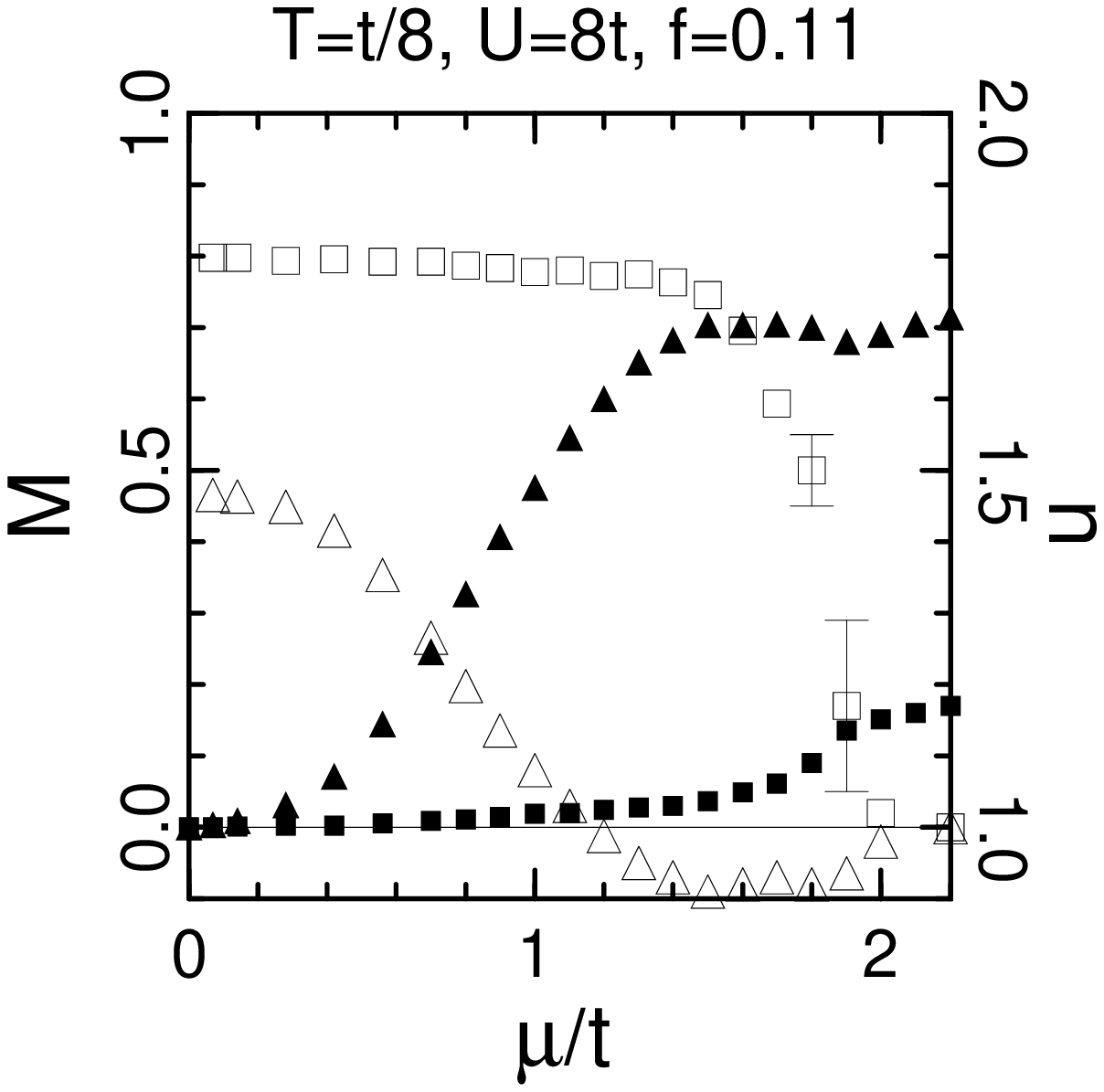,width=4.in,angle=0}}
\vspace{-10mm}
\caption{
Density $n$ (solid symbols) and sublattice magnetization $M$ (open symbols)
vs.~$\mu$ in DMFT separately for $U=0$ sites 
(triangles) and $U=8t$ sites (squares).
For small $\mu$ there is AF order on both types of sites. For $\mu>1.6t$
the density on the $U=8t$ sites increases and AF order breaks down.
\label{nvsmudinf}}
\vspace*{-122mm}
\hspace*{65mm}
{\psfig{file=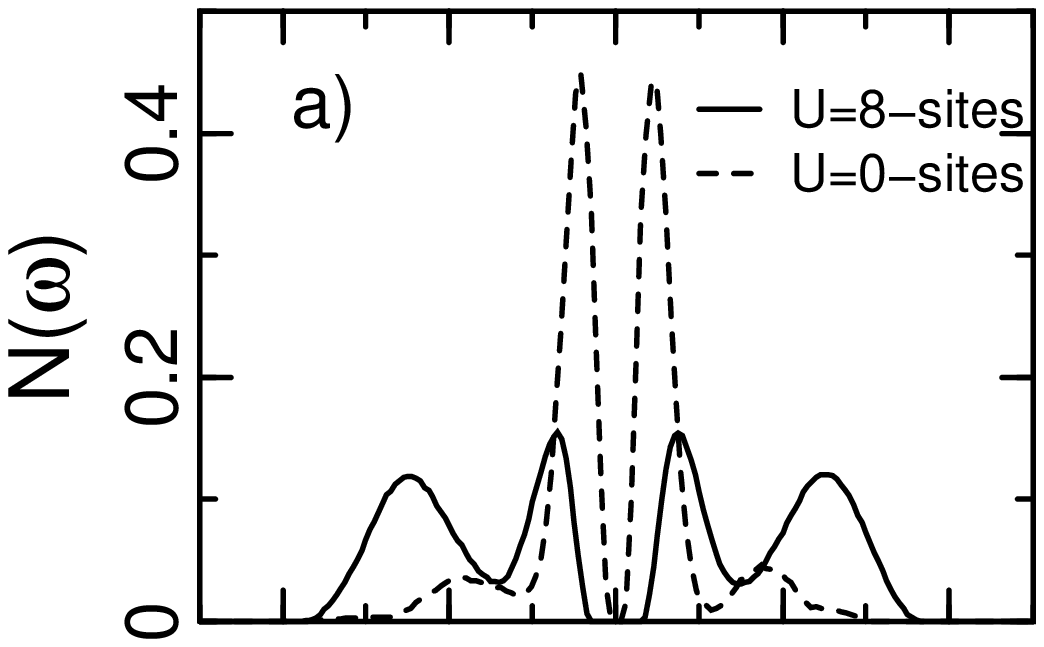,width=4.in,height=2.3in,angle=0}}
\hspace*{65mm}
{\psfig{file=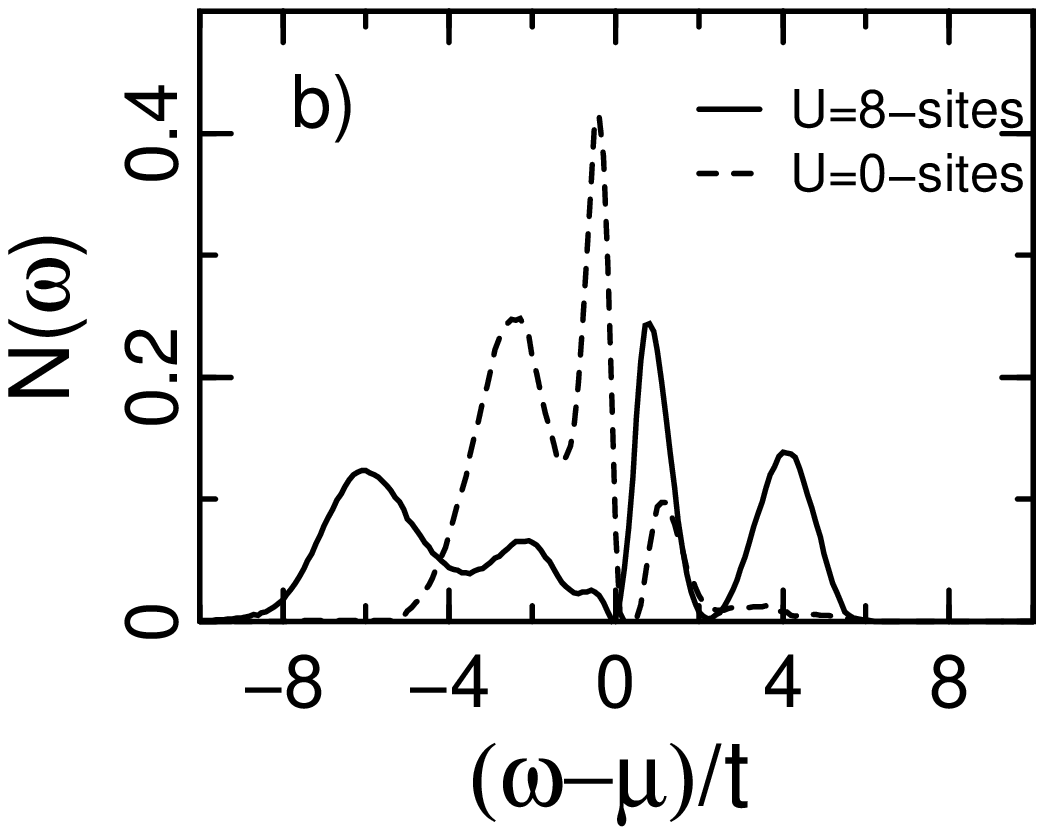,width=4.in,height=2.3in,angle=0}}
\caption{
Density of states in DMFT on $U=0$ and $U=8t$ sites for $T=t/8$, $f=0.11$
at a) half--filling and b) $n=1.11$. All spectra have a vanishing DOS
at the Fermi energy.
\label{DOS}}
\end{figure}

In order to elucidate the nature of this remnant gap at half--filling,
we examine the effect of the chemical potential on the
staggered magnetization $M$. 
Fig.~\ref{nvsmudinf} shows DMFT results for the densities and sublattice 
magnetizations vs.~$\mu$ at $f=0.11$ for $U=0$ and $U=8t$ sites separately. 
The densities show quantitatively the same dependence 
on $\mu$ in $d=2$ and in DMFT \cite{Denteneer98},
however the sublattice magnetization off half--filling can be obtained
in DMFT only. 
The density on the $U=8t$ sites hardly changes for $\mu<1.6t$,
but then begins to rise, at which point $M$
also abruptly vanishes. 
For small $\mu$ there is induced AF order also on the $U=0$ sites.
This AF order leads to a gap even on the $U=0$ sites due to
the doubling of the unit cell.
%
The two different gap values on the $U=8t$ and $U=0$ sites
explain the double gap structure in the total density (Fig.~\ref{nvsmud2}).

Surprisingly, $M$ becomes
\em negative \em on the $U=0$ sites when $n(U=0)$ starts to saturate.
The reason is that an electron on a $U=0$ site with spin parallel to
its neighbors is more strongly localized (due to Pauli's principle)
than an electron with opposite spin. Thus the net moment on the $U=0$ site 
is \em parallel \em to its neighbors, i.e.~opposite to the total staggered
magnetization.  Within the DMFT approach AF order is 
more stable against disorder ($f_c^{\infty}=0.75\pm 0.02$ at $T=0$ and $n=1$, 
compared to $f_c\approx 0.4$ in $d=2$), as is typical for mean--field theories.

The charge gap can also be obtained from the density of states (DOS).
We employ the DMFT
approach, combined with the ``Maximum Entropy'' method \cite{Jarrell96},
which gives excellent agreement with $d=2$ and
$d=3$ results in the clean case \cite{Bulut94,Preuss95}.
At half--filling and low $T$ (Fig.~\ref{DOS}a) the DOS for the $U=8t$ sites
shows the typical AF structure \cite{Preuss95}
with two broad Hubbard bands, two quasi--particle peaks at low energies,
and a gap at the Fermi energy, $E_F=0$. 
On the $U=0$ sites the Hubbard bands appear at lower energies and 
the gap is very small but still present, and will increase at lower $T$. 
Thus there is indeed a gap in the antiferromagnetic phase,
even for the $U=0$ sites. 
For small doping $(1.0<n<1.11)$ the compressibility was seen
to be large (see Figs.~\ref{nvsmud2} and \ref{nvsmudinf}).
At $n=1.11$, where the density on the $U=0$ sites saturates,
the DOS again vanishes at the Fermi energy, $E_F=\mu=1.45t$
(Fig.~\ref{DOS}b). At this point there is still AF order.

\vspace*{1em}
\em Summary. \em ---
We can summarize the physics of the charge gap and the AF order
in terms of the phase diagram of Fig.~\ref{phased}.
In the pure case AF order breaks down when the density  
reaches a critical value $n_c(0)$. This value
depends on dimensionality and may very well be 1.0 in $d=2$.
At a finite concentration $f$, the defect sites localize dopants,
thereby keeping the rest of lattice at half filling, 
until they are (almost) doubly occupied.  
For small concentration $f$ the defect sites hence 
stabilize AF order up to a total density of 
$n_c(f) \approx n_c(0) (1-f) + 2 f$.
Eventually, at large enough $f$, $n_c$ is driven to one and
AF order ceases to exist even for $n=1$.
The density exhibits a plateau associated
with AF order, which exists on both the nonzero $U$
and the $U=0$ sites, as the chemical
potential is changed across half--filling ($n=1$);
and a second ``Mott'' plateau, associated with double occupancy, 
as one crosses the high filling boundary of the AF region.  

\vspace*{1em}
We thank Phil Rogers for discussions on Zn doping
of high temperature superconductors.
This work was supported by NSF--DMR--95--28535 and the SDSC.   


\begin{thebibliography}{99}


\bibitem{GEBHARD}
For a recent review see: 
F.~Gebhard, {\em The Mott Metal--Insulator Transition}, 
Sprin\-ger Tracts in Modern Physics, vol.~137 (Springer, Heidelberg, 1997).

\bibitem{ZN1}
G.~Xiao, M.~Z.~Cieplak, A.~Gavrin, F.~H.~Streitz, A.~Bakhshai, and C.~L.~Chien
Phys.~Rev.~Lett.~{\bf 60}, 1446 (1988);
B.~Keimer, A.~Aharony, A.~Auerbach, R.~J.~Birgeneau, A.~Cassanho, Y.~Endoh,
R.~W.~Erwin, M.~A.~Kastner, G.~Shirane, Phys.~Rev.~B {\bf 45}, 7430 (1992);
A.~V.~Mahajan, H.~Alloul, G.~Collin, and J.~F.~Marucco,
Phys.~Rev.~Lett.~{\bf 72}, 3100 (1994).

\bibitem{ZN2}
M.~Azuma, Y.~Fujishiro, M.~Takano, 
M.~Nohara, and H.~Takagi, Phys.~Rev.~B {\bf 55}, 8658 (1997). 

\bibitem{NUMERICS}
G.~B.~Martins, E.~Dagotto, and J.~Riera, 
Phys.~Rev.~B {\bf 54}, 16032 (1996); Y.~Motome, N.~Katoh, N.~Furukawa, 
and M.~Imada, j.~Phys.~Soc.~Jpn.~{\bf 65}, 1949 (1996).
S.~Miyashita and S.~Yamamoto,
Phys.~Rev.~B {\bf 48}, 913 (1993); 
E.~Sorensen and I.~Affleck,
Phys.~Rev.~B {\bf 49}, 15771 (1994); 
N.~Bulut, D.J.~Scalapino, and E.~Loh, 
Phys.~Rev.~Lett.~{\bf 62}, 2192 (1989).

\bibitem{MARTINS}
G.~B.~Martins, M.~Laukamp, J.~Riera, and
E.~Dagotto, Phys.~Rev.~Lett.~{\bf 78}, 3563 (1997).

\bibitem{Blankenbecler81} R.~Blankenbecler, D.~J.~Scalapino, and R.~L.~Sugar,
 Phys.~Rev.~D {\bf 24}, 2278 (1981); 
S.~R.~White, D.~J. Scalapino, R.~L.~Sugar, E.~Y.~Loh,
J.~E.~Gubernatis, R.~T.~Scalettar, Phys.~Rev.~B {\bf 40}, 506 (1989).

\bibitem{Sugiyama86etc}
G.~Sugiyama and S.~E.~Koonin, Ann.~Phys.~{\bf 168}, 1 (1986);
S.~Sorella, S.~Baroni, R.~Car, and M.~Parrinello, 
Europhys.~Lett.~{\bf 8}, 663 (1989).

\bibitem{Metzner89etc}
W.~Metzner and D.~Vollhardt, Phys.~Rev.~Lett.~{\bf 62}, 324 (1989); for a
review see D.~Vollhardt, in {\em Correlated Electron Systems}, edited by V.
J.~Emery (World Scientific, Singapore, 1993), p.~57.
\bibitem{Jarrell92}
M.~Jarrell, Phys.~Rev.~Lett.~{\bf 69}, 168 (1992);
T.~Pruschke, M.~Jarrell, and J.~K. Freericks,
Adv.~Phys. {\bf 44}, 187 (1995). 

\bibitem{Georges96}
A.~Georges, G.~Kotliar, W.~Krauth, and M.~Rozenberg,
Rev.~Mod.~Phys.~{\bf 68}, 13 (1996).

\bibitem{Hirsch86} J.~E.~Hirsch and R.~M.~Fye,
Phys.~Rev.~Lett.~{\bf 56}, 2521 (1986).

\bibitem{Dobrosalievic93etc}
V.~Dobrosavljevi\'{c} and G.~Kotliar,
Phys.~Rev.~Lett.~{\bf 71}, 3218 (1993); 
Phys.~Rev.~B {\bf 50}, 1430 (1994).

\bibitem{Ulmke95} M.~Ulmke, V.~Jani\v s, and D.~Vollhardt, 
Phys.~Rev.~B {\bf 51}, 10411 (1995).

\bibitem{Elliot74} R.~J.~Elliot, J.~A.~Krumhansl, and P.~L.~Leath,
Rev. Mod.~Phys.~{\bf 46}, 465 (1974).

\bibitem{phsymmetry} Particle-hole symmetry is also preserved in the case of 
random nearest--neighbor hopping. 
While random potentials break the particle-hole symmetry
this does not necessarily imply a strong suppression of AF order, 
see e.g.~\cite{Ulmke95}.

\bibitem{Huse88} D.~A.~Huse, Phys.~Rev.~B {\bf 37}, 2380 (1988).

\bibitem{ERRORBARS} We generate
average values and error bars using the ``midmean''
to deal with large sign problem fluctuations;
J.~L.~Rosenberger and M.~Gasko, Ch.~10 in {\em Understanding Robust and
Exploratory Data Analysis}, edited by D.C.~Hoaglin, F.~Mosteller, and
J.~W.~Tukey (Wiley, New York, 1983). More details are given in 
\cite{Denteneer98}.

\bibitem{Denteneer98} P.~J.~H.~Denteneer, M.~Ulmke, R.~T.~Scalettar,  
and G.~T.~Zimanyi, Physica A {\bf 251}, 162 (1998).

\bibitem{CPA}
The enhancement of $M$ with $f$ is found to be absent in the dynamical
mean--field theory \cite{Denteneer98}.
The reason is that the spin wave reduction of $M$ vanishes 
in high dimensions.

\bibitem{Bulut94} N.~Bulut, D.~J.~Scalapino, and S.~R.~White, 
Phys.~Rev.~Lett.~{\bf 72}, 705 (1994); Phys.~Rev.~B {\bf 50}, 7215 (1994);
Phys.~Rev.~Lett.~{\bf 73}, 748 (1994).

\bibitem{Preuss95} R.~Preuss, W.~Hanke, and W.~von der Linden, 
Phys.~Rev.~Lett.~{\bf 75}, 1344 (1995). 
M.~Ulmke, R.~T.~Scalettar, A.~Nazarenko, 
and E. Dagotto, Phys.~Rev.~{\bf B 54}, 16523 (1996).

\bibitem{Jarrell96}  For a review see M.~Jarrell and J.~E.~Gubernatis, 
Phys.~Rep.~{\bf 269}, 133 (1996).

\end{thebibliography}
\end{document}